\documentclass[12pt]{article}
\pdfoutput=1

\usepackage{amssymb}
\usepackage{amsmath}
\usepackage{amscd}
\usepackage{mathrsfs}
\usepackage{amssymb}
\usepackage{feynmf}
\usepackage[pdftex]{graphicx}

\def\dj{\hbox{d\kern-0.347em \vrule width 0.3em height 1.252ex depth
-1.21ex \kern 0.051em}}

\numberwithin{equation}{section}

\begin{document}

\setlength{\oddsidemargin}{0cm}
\setlength{\baselineskip}{7mm}


\thispagestyle{empty}
\setcounter{page}{0}

\begin{flushright}

\end{flushright}

\vspace*{1cm}

\begin{center}
{\bf \Large Galileon Radiation from a Spherical Collapsing Shell}

\vspace*{1cm}

Javier Mart\'{\i}n-Garc\'{\i}a$^{a,}\footnote{\tt 
javoxmg@gmail.com}$
and Miguel \'A. V\'azquez-Mozo$^{b,}$\footnote{\tt 
Miguel.Vazquez-Mozo@cern.ch}

\end{center}

\vspace*{0.0cm}

\begin{center}

$^{a}${\sl Instituto de F\'{\i}sica Te\'orica UAM/CSIC \\
C/ Nicol\'as Cabrera 15, E-28049 Madrid, Spain
}

$^{b}${\sl Instituto Universitario de F\'{\i}sica Fundamental y Matem\'aticas (IUFFyM) \\
 Universidad de Salamanca \\ 
 Plaza de la Merced s/n,
 E-37008 Salamanca, Spain
  }
\end{center}

\vspace*{2.5cm}

\centerline{\bf \large Abstract}

\noindent
Galileon radiation in the collapse of a thin spherical shell of matter is analyzed. In the framework 
of a cubic Galileon theory,  
we compute the field profile produced at large distances by a short collapse,
finding that the radiated field has two peaks traveling ahead of light fronts.  
The total energy radiated during the collapse follows a
power law scaling with the shell's physical width and results from two competing 
effects: a Vainshtein suppression of the emission and an enhancement due to the thinness of the shell.

\newpage

\section{Introduction}

The discovery of the accelerated expansion of the universe \cite{supernovae} triggered a variety of attempts to modify the infrared sector
of gravity in order to avoid the introduction of a cosmological constant, whose value cannot be explained naturally 
in the framework of quantum field theory~\cite{CC}. Among the earlier proposals was the five-dimensional 
Dvali-Gabadadze-Porrati (DGP) model~\cite{DGP}, in which the gravitational dynamics is governed by an action containing two 
Einstein-Hilbert terms for both the ambient metric and its pullback on our four-dimensional braneworld. The hierarchy between
the five- and four-dimensional Planck masses results 
in an effective mass term for the four-dimensional graviton. Its phenomenological prospects have been widely studied
(see, e.g., \cite{lue} and references therein).

One consequence of the formulation of the DGP model was a renewed interest in massive gravity, 
which has turned out to be one of the 
the most interesting large scale modifications of gravity studied in the last decade (see \cite{hinterbichlerRMP,deRhamLRR} for reviews). 
The graviton mass breaks general covariance, which allows in principle for a degravitation 
of the cosmological constant term \cite{degravitation}. 
Invariance under coordinate transformations can nevertheless be restored through the
introduction of St\"uckelberg fields \cite{stueckelberg}. 

Massive gravity, at least in its most naive formulation, is not free from potential problems. One of them is that  
the additional graviton polarizations do not decouple in the limit of zero mass, so General Relativity (GR) is not recovered. 
This van Dam-Veltman-Zakharov (vDVZ) 
discontinuity is overcome through the Vainshtein mechanism \cite{vainshtein}, in that the strong nonlinearities of the massive
gravity Lagrangian screen the matter coupling of the massive graviton's scalar mode at distances below some 
characteristic Vainshtein radius. 
A second problem is the emergence of classical Ostrogradsky instabilities \cite{ostrogradsky} or ghost states at the quantum level, 
in particular the so-called Boulware-Deser ghost
\cite{ghost1}. This dangerous mode is avoided through a resummation of nonlinear terms \cite{dRGT_ghostfree}, leading a ghost-free theory 
of massive gravity at all orders.

In both the DGP model and massive gravity, there is a limit in which the massive graviton's scalar mode $\pi(x)$ decouples from the 
transverse components $h_{\mu\nu}(x)$, resulting in a scalar field theory invariant under Galilean transformations,
$\pi(x)\rightarrow \pi(x)+a+b_{\mu}x^{\mu}$, and characterized by an energy scale 
\begin{eqnarray}
\Lambda=(M_{\rm Pl}m^{2})^{1\over 3},
\label{eq:lambda_scale_def}
\end{eqnarray} 
with $m$ the graviton mass.
These Galileon theories \cite{galileon} have some interesting properties as 
field theories: the scale $\Lambda$ is stable under quantum corrections
and there is a regime in which non-Galileon interactions remain unimportant \cite{nicolis_rattazzi,galileon}.  
Galileon field theory has been extensively studied in a number of physical setups \cite{deRhamLRR,galileon_reviews}. 

Gravitational collapse is a powerful testbench in gravitational physics. Analyzing the problem of a 
collapsing sphere of dust, Oppenheimer and Snyder \cite{oppenheimer-snyder} were able to glipmse the 
nonsingular character of the horizon, decades before a mathematical solution to the issue was 
available. As an ubiquitous process in astrophysics, it is the source of many observational
signals in the Universe \cite{grav_collapse_revs}. 

There are several reasons justifying the study of gravitational collapse in the context of massive gravity and Galileon theories.
In GR, Birkhoff's theorem prevents the emission of gravitational radiation from spherical collapse. 
Gravitational theories with scalar degrees of freedom, on the other hand, allow the radiation of energy 
even when spherical symmetry is preserved \cite{sexl}. 
The opening of new channels for the radiation of energy can be relevant in a number of astrophysical processes and 
might be used to put the theory to the test. 
In the case of massive generalizations of GR, and particularly in Galileon theories, the very special features of
these scalar modes might lead to distinct observational signals.

So far, gravitational collapse in Galileon theories has been studied mostly in the
context of structure formation \cite{structure_form_papers} and the Vaidya solution \cite{vaidya_solution}.
In this paper we analyze the problem of Galileon emission at the onset of the gravitational collapse of a 
spherical thin shell of matter. Our model consists in a delta-function shell that starts collapsing with or without 
initial velocity, stoping collapse after a short time. 
Due to its coupling through the trace of the energy-momentum tensor, this collapsing matter 
introduces a time-dependent perturbation acting as a source for a radiating Galileon field. 

One of the consequences of considering the ideal situation of a delta-function shell is that we have field gradients 
above the Galileon scale $\Lambda$, leading to breakdown of effective field theory. In addition, the total energy radiated during
the process diverges due to the contribution of arbitrary high frequencies. In order to avoid these problems we carry out our
calculations using a physical cutoff in frequencies whose value is determined by the physical width of the shell, which 
we take to be much larger than the cutoff scale $\Lambda^{-1}$. What we find is that 
the profile of the Galileon field detected at large distances exhibits two pulses propagating
ahead of light fronts. As for the total energy radiated, we obtain a very simple scaling with the shell's physical width. 

The plan of the paper is as follows. In the next section we present the model to be studied, an imploding delta-function
spherical shell collapsing under its own gravity. In Section \ref{sec:pert} we detail the perturbative
approach to be used and solve for the profile of the Galileon field at large distances from the source. After this, 
the total energy radiated is computed in Section \ref{sec:energy}, whereas Section \ref{sec:nlo} is devoted to the
analysis of the next-to-leading order correction, and in particular to the case in which the collapse starts from rest. 
Finally, in Section \ref{eq:conc} we comment on some possible directions for future work. 

\section{The model}
\label{sec:model}

We work in the context of a cubic Galileon theory with Lagrangian \cite{deRhamLRR}
\begin{eqnarray}
\mathcal{L}=-{1\over 2}\partial_{\mu}\pi\partial^{\mu}\pi-{1\over \Lambda^{3}}\partial_{\mu}\pi\partial^{\mu}\pi\Box\pi+
{1\over M_{\rm Pl}}\pi T,
\label{eq:cubic?galileon}
\end{eqnarray}
where $\Lambda$ is the Galileon energy scale, $M_{\rm Pl}$ is the Planck mass, and $T$ is the trace of the matter energy-momentum tensor. 
To address the problem of Galileon emission in the gravitational collapse of a spherical source we consider the 
following form of the
energy-momentum tensor
\begin{eqnarray}
T_{\mu\nu}=\rho(t,r)\delta^{0}_{\mu}\delta^{0}_{\nu}.
\label{eq:em-tensor-generic}
\end{eqnarray}
The time evolution of $\rho(t,r)$ is determined by the 
equations of gravity. 
In our calculation we also follow the strategy of \cite{dRMT1} and consider that time
evolution is treated as a perturbation on a static background. In other words, 
we treat the problem perturbatively and split the energy-momentum tensor into a
static background and a dynamic perturbation
\begin{eqnarray}
\rho(t,r)=\rho_{0}(r)+\delta\rho(t,r),
\label{eq:split_back_pert_dens}
\end{eqnarray}
while the Galileon field is also split accordingly as
\begin{eqnarray}
\pi(t,r)=\pi_{0}(r)+\phi(t,r),
\label{eq:split_galileon_field}
\end{eqnarray}
where $\pi_{0}(r)$ is a static, spherically symmetric solution to the Galileon field equations \cite{spherical_galileons}.

To be more specific, let us focus on the Galileon equations sourced by an energy-momentum tensor associated with a static spherical shell 
located a the
position $r=R_{0}$,
\begin{eqnarray}
\rho_{0}(r)=\sigma_{0}\delta(r-R_{0}),
\end{eqnarray}
where $\sigma_{0}$ is the superficial density of the shell. In choosing a spherical shell instead
of a ball we simplify the analysis in that Galileons are emitted only at the surface of the
collapsing body and not from the interior, which would be the case if $\delta T\neq 0$ for 
$r<R$. This can be seen as a rough model of the collapse of an outer layer of an astrophysical object over its core.

The equations of motion for the background helicity-0 mode in the cubic Galileon theory \eqref{eq:cubic?galileon} have the form
\cite{chu_trodden}
\begin{eqnarray}
{1\over r^{2}}\partial_{r}\left\{r^{3}\left[{\pi_{0}'\over r}+{4\over\Lambda^{3}}\left({\pi_{0}'\over r}\right)^{2}\right]\right\}
={\sigma_{0}\over M_{\rm Pl}}\delta(r-R_{0})
.
\label{eq:back3galileon}
\end{eqnarray}
To solve them, we look for solutions outside and inside the shell and  
match them across $r=R$, using the conditions derived from integrating Eq. \eqref{eq:back3galileon},
\begin{eqnarray}
\int_{R_{0}-\epsilon}^{R_{0}+\epsilon}dr\,
\partial_{r}\left\{r^{3}\left[{\pi_{0}'\over r}+{4\over\Lambda^{3}}\left({\pi_{0}'\over r}\right)^{2}\right]\right\}
={\sigma_{0} R_{0}^{2}\over M_{\rm Pl}},
\end{eqnarray}
which gives
\begin{eqnarray}
\left.r^{3}\left[{\pi_{0}'\over r}+{4\over\Lambda^{3}}\left({\pi_{0}'\over r}\right)^{2}\right]\right|_{R_{0}+\epsilon}
-\left.r^{3}\left[{\pi_{0}'\over r}+{4\over\Lambda^{3}}\left({\pi_{0}'\over r}\right)^{2}\right]\right|_{R_{0}-\epsilon}
={\sigma_{0} R_{0}^{2}\over M_{\rm Pl}}.
\label{eq:cubicG_matching_cond}
\end{eqnarray}

With this result, we integrate Eq. \eqref{eq:back3galileon} over a ball of radius $r>R_{0}$, 
obtaining
\begin{eqnarray}
{\pi_{0}'\over r}+{4\over\Lambda^{3}}\left({\pi_{0}'\over r}\right)^{2}={\sigma_{0} R_{0}^{2}\over M_{\rm Pl}r^{3}}
\hspace*{1cm} (r>R_{0}).
\label{eq:ext_shell_cubicG}
\end{eqnarray}
This gives a quadratic equation for ${\pi_{0}'/r}$ whose solutions are 
\begin{eqnarray}
{\pi_{0}'\over r}=-{\Lambda^{3}\over 8}\left(1\mp \sqrt{1+{16\sigma_{0} R_{0}^{2}\over \Lambda^{3}M_{\rm Pl}r^{3}}}\right)
\hspace*{1cm} (r>R_{0}).
\end{eqnarray}
The condition that the Galileon vanishes at infinity selects the $-$ branch. Notice that this is the same solution than
for the case of a pointlike particle with mass $m=4\pi\sigma R_{0}^{2}$. 

For the shell interior, we just integrate the homogeneous equation
\begin{eqnarray}
{1\over r^{2}}\partial_{r}\left\{r^{3}\left[{\pi_{0}'\over r}+{4\over\Lambda^{3}}\left({\pi_{0}'\over r}\right)^{2}\right]\right\}
=0,
\end{eqnarray}
with the result
\begin{eqnarray}
{\pi_{0}'\over r}+{4\over\Lambda^{3}}\left({\pi_{0}'\over r}\right)^{2}={C\over r^{3}} \hspace*{1cm}
(r<R_{0}).
\label{eq:int_shell_cubicG}
\end{eqnarray}
Plugging Eqs. \eqref{eq:ext_shell_cubicG} and \eqref{eq:int_shell_cubicG} into the matching condition
\eqref{eq:cubicG_matching_cond}, we fix the value of the integration constant $C$ to be 
\begin{eqnarray}
{\sigma_{0} R_{0}^{2}\over M_{\rm Pl}}-C={\sigma_{0} R_{0}^{2}\over M_{\rm Pl}} \hspace*{1cm} \Longrightarrow \hspace*{1cm}
C=0.
\end{eqnarray}
Equation \eqref{eq:int_shell_cubicG} has therefore two solutions: a trivial one $\pi'_{0}=0$ together with
\begin{eqnarray}
{\pi_{0}'\over r}=-{\Lambda^{3}\over 4} \hspace*{1cm} (r<R_{0}).
\end{eqnarray}
 
To find the right background solution for the cubic Galileon we have to take into account that for $\sigma_{0}\rightarrow 0$ we should
recover a continuous ``vacuum'' solution $\pi'_{0}=0$. Thus, we are forced to choose the trivial solution for the interior of the shell and write
\begin{eqnarray}
{\pi_{0}'(r)\over r}=-{\Lambda^{3}\over 8}\left(1- \sqrt{1+{16\sigma_{0} R_{0}^{2}\over M_{\rm Pl}\Lambda^{3} r^{3}}}\right)
\theta(r-R_{0}), 
\label{eq:solution_pi0'}
\end{eqnarray}
where $\theta(x)$ is the Heaviside step function. 
We read the value of the Vainshtein radius off this expression, with the result
\begin{eqnarray}
r_{\star}=\left({16\sigma_{0} R_{0}^{2}\over M_{\rm Pl}\Lambda^{3}}\right)^{1\over 3}.
\end{eqnarray}
The solution for the background Galileon field $\pi_{0}(r)$ obtained by integrating Eq. \eqref{eq:solution_pi0'} is continuous. The
discontinuity in its first radial derivative at $r=R_{0}$ is a consequence of the field being sourced by an infinitely 
thin distribution of matter. 

Thus, our problem has two natural length scales: the radius of the shell $R_{0}$ and the Vainshtein radius $r_{\star}$. 
Let us asume first
that the Vainshtein radius is (much) smaller than 
the radius of the shell. 
To see whether this approximation is physically relevant, we rewrite the condition $R_{0}\gg r_{\star}$ as
\begin{eqnarray}
\sigma_{0} \ll {1\over 16}M_{\rm Pl}\Lambda^{3}R_{0},
\end{eqnarray}
which can be recast as 
\begin{eqnarray}
4\pi_{0} R_{0}^{2}\sigma_{0} \ll {\pi\over 4} M_{\rm Pl}\Lambda^{3}R_{0}^{3}={3\over 16}M_{\rm Pl}\Lambda^{3}V_{0},
\end{eqnarray}
with $V_{0}$ the volume enclosed by the shell. Defining the equivalent density of the shell as
\begin{eqnarray}
\rho_{\rm equiv}\equiv {4\pi R_{0}^{2}\sigma_{0}\over V_{0}},
\end{eqnarray}
we get the bound
\begin{eqnarray}
\rho_{\rm equiv}\ll {3\over 16}M_{\rm Pl}\Lambda^{3}.
\end{eqnarray}
We take the usual value \cite{hinterbichlerRMP} for the cutoff scale $\Lambda\simeq (1000{\rm \,km})^{-1}$, which is obtained 
from \eqref{eq:lambda_scale_def} by assuming a graviton mass of the order of the Hubble scale, 
$m\sim H_{0}^{-1}$.
This value is around the current bounds for the graviton mass \cite{graviton_mass_bounds}. 
With this we arrive at
\begin{eqnarray}
\rho_{\rm equiv}\ll 10^{-47} \,\,{\rm GeV}^{4},
\end{eqnarray}
where the bound is of the order of the present energy density of the universe. 
This energy density is completely negligible in an astrophysical setup, so 
in order to have a physically meaningful model we exclude the
case when the Vainshtein radius is much smaller than the radius of the shell. In the following we will work in the case where 
the radius of the shell lies well inside the Vainshtein radius, $R_{0}\ll r_{\star}$. 

\section{Perturbative analysis}
\label{sec:pert}

Inserting the decomposition \eqref{eq:split_galileon_field} into the Lagrangian for the cubic Galileon theory 
\eqref{eq:cubic?galileon} and keeping terms quadratic in the perturbed quantities leads to the following
Lagrangian for the perturbation in the Galileon field $\phi(x)$
\begin{eqnarray}
\mathcal{L}_{\rm pert}=-{1\over 2}Z^{\mu\nu}\partial_{\mu}\phi\partial_{\nu}\phi+{1\over M_{\rm Pl}}\phi\,\delta T,
\end{eqnarray}
where the effective metric $Z_{\mu\nu}$ is given by \cite{deRhamLRR}
\begin{eqnarray}
Z_{\mu\nu}dx^{\mu}dx^{\nu}&=& -\left[1+{4\over \Lambda^{3}}\left({2\pi_{0}'\over r}+\pi_{0}''\right)\right]dt^{2}
+\left(1+{8\pi_{0}'\over r\Lambda^{3}}\right)dr^{2}\nonumber \\
&+&\left[1+{4\over \Lambda^{3}}\left({\pi_{0}'\over r}+
\pi_{0}''\right)\right]r^{2}d\Omega^{2}.
\label{eq:effective_metric_pert}
\end{eqnarray}
In terms of this, the equations of motion read
\begin{eqnarray}
\Box_{Z}\phi\equiv \partial_{\mu}(Z^{\mu\nu}\partial_{\nu}\phi)=-{1\over M_{\rm Pl}}\delta T.
\label{eq:laplacian_eff_metric}
\end{eqnarray}

The radiating Galileon field is sourced by the perturbation in the trace of the energy-momentum tensor. In the case of a collapsing
shell, we have
\begin{eqnarray}
T=-\sigma(\tau)\delta\Big(r-R(\tau)\Big),
\label{eq:trace_em_tensor}
\end{eqnarray}
where $\tau$ is the proper time for an observer falling with the shell and $R(\tau)$ is given by the solution to the equation
\cite{poisson}
\begin{eqnarray}
M=4\pi R_{0}^{2}\sigma_{0}\sqrt{\dot{R}(\tau)^{2}+1}-{8\pi^{2}G_{N}R_{0}^{4}\sigma_{0}^{2}\over R(\tau)}.
\end{eqnarray}
Here, $M$ is the mass of the shell as seen by a distant observer and $R_{0}$, $\sigma_{0}$ are the initial values of the radius and surface energy density respectively. The first term on the right-hand side of this equation can be interpreted as the kinetic energy of the shell, whereas the second one is its gravitational binding energy. 
Once $R(\tau)$ is found, the (exterior) time coordinate at the location of the shell is given in terms of proper time by the solution to the
equation
\begin{eqnarray}
\dot{t}(\tau)=\left[1-{2G_{N}M\over R(\tau)}\right]^{-1}\sqrt{\dot{R}(\tau)^{2}+1-{2G_{N}M\over R(\tau)}}.
\end{eqnarray}
Finally, the time evolution of the surface density
is given in terms of $R(\tau)$ by
\begin{eqnarray}
\sigma(\tau)=\sigma_{0}\left[{R_{0}\over R(\tau)}\right]^{2}.
\label{eq:sigma(tau)}
\end{eqnarray}

Let us consider a physical situation in which the shell is stable for negative times  
\begin{eqnarray}
R(\tau)=R_{0} \hspace*{1cm} \mbox{for} \hspace*{1cm} \tau<0.
\end{eqnarray}
and that at $\tau=0$ it implodes with initial velocity $\dot{R}_{0}$ during a short proper time $\delta\tau$.  
The corresponding perturbation on the static energy-momentum tensor \eqref{eq:em-tensor-generic} 
induced by time evolution is given by
\begin{eqnarray}
\delta T=-\dot{\sigma}(\tau)\delta \tau\delta\Big(r-R(\tau)\Big)+\sigma(\tau)\dot{R}(\tau)\delta\tau\delta'\Big(r-R(\tau)\Big).
\label{eq:deltaT}
\end{eqnarray}
while $\delta T=0$ for $\tau<0$ and $\tau>\delta\tau$.
Using the equation for the time evolution of the surface energy density \eqref{eq:sigma(tau)} we can eliminate $\dot{\sigma}(\tau)$
to write
\begin{eqnarray}
\delta T={\delta\tau \dot{R}(\tau)\over R(\tau)}\sigma(\tau)\left[R(\tau)\delta'\Big(r-R(\tau)\Big)+2
\delta\Big(r-R(\tau)\Big)\right],
\label{eq:form_of_perturbation}
\end{eqnarray}
where our expansion parameter is 
\begin{eqnarray}
{\delta\tau \dot{R}(\tau)\over R(\tau)}\ll 1.
\end{eqnarray}

Once $\delta T$ is known, the corresponding perturbation in the Galileon field $\phi(t,r)$ can
be computed as
\begin{eqnarray}
\phi(x)={1\over M_{\rm Pl}}\int d^{4}x'\,G(x,x')\delta T(x'),
\label{eq:phi_solution}
\end{eqnarray}
where $G(x,x')$ is the retarded Green function of the Laplacian operator defined in Eq. \eqref{eq:laplacian_eff_metric}. 
This object has been studied in \cite{chu_trodden}.
In spherical coordinates, it is explicitly given by the solution to the equation
\begin{eqnarray}
\left[e_{1}(r)\partial_{t}^{2}-e_{2}(r)\partial_{r}^{2}-{2\over r}e_{3}(r)\partial_{r}
-{e_{3}(r)\over r^{2}}\mathbf{L}^{2}\right]
G(t,r,\varphi,\theta;t',r',\varphi',\theta')\hspace*{3cm} \nonumber \\[0.2cm]
={1\over rr'}\delta(t-t')\delta(r-r')
\delta(\varphi-\varphi')\delta(\cos{\theta}-\cos{\theta'}),
\end{eqnarray}
where $\mathbf{L}^{2}$ denotes the Laplacian over the transverse two-dimensional unit sphere 
and the functions $e_{i}(r)$ are given by
\begin{eqnarray}
e_{1}(r)&=&{3\over 4}\left[{2r^{3}+r_{\star}^{3}\over \sqrt{r^{3}(r^{3}+r_{\star}^{3})}}-{2\over 3}
\right], \nonumber \\[0.2cm]
e_{2}(r)&=&\sqrt{1+{r_{\star}^{3}\over r^{3}}}, \\[0.2cm]
e_{3}(r)&=&{4r^{3}+r_{\star}^{3}\over 4\sqrt{r^{3}(r^{3}+r_{\star}^{3})}}. \nonumber
\end{eqnarray}

Using the fact that the coefficients are time independent, the
Green function can be expanded as \cite{chu_trodden}
\begin{eqnarray}
G(x,x')=\int_{-\infty}^{\infty}{\omega d\omega\over 2\pi} e^{-i\omega(t-t')}
\sum_{\ell=0}^{\infty}\widetilde{g}_{\ell}(\omega r,\omega r')
\sum_{m=-\ell}^{\ell}Y^{m}_{\ell}(\theta,\varphi)\overline{Y}^{m}_{\ell}(\theta',\varphi'),
\label{eq:Green_func_full}
\end{eqnarray}
where the radial part of the Green function satisfies the equation
\begin{eqnarray}
\left[e_{2}(\xi)\partial_{\xi}^{2}+{2\over \xi}e_{3}(\xi)\partial_{\xi}+e_{1}(\xi)-
{\ell(\ell+1)\over \xi^{2}}e_{3}(\xi)\right]\widetilde{g}_{\ell}(\xi,\xi')=
{1\over \xi\xi'}\delta(\xi-\xi').
\end{eqnarray}
Due to the spherical symmetry of the gravitational collapse under study, 
the multipole expansion of the Green function gets truncated to the monopole term
$\ell=0$. This means that Eq. \eqref{eq:phi_solution} reads
\begin{eqnarray}
\phi(x)&=&{1\over 4\pi M_{\rm Pl}}\int_{-\infty}^{\infty}{\omega d\omega\over 2\pi}\int d^{4}x'e^{-i\omega(t-t')}
\widetilde{g}_{0}(\omega r,\omega r')\delta T(x') \\[0.2cm]
&=&{1\over M_{\rm Pl}}\int_{-\infty}^{\infty}{\omega d\omega\over 2\pi}e^{-i\omega t}
\int_{0}^{\delta\tau} d\tau\,{\sqrt{\dot{R}(\tau)^{2}+f(\tau)}\over f(\tau)}e^{i\omega\tau}\int_{0}^{\infty}r'^{2}dr'\,
\widetilde{g}_{0}(\omega r,\omega r')\delta T(\tau,r'), 
\nonumber
\end{eqnarray}
where in the second line we have exploited the fact that the perturbation to the energy-momentum tensor is spherically 
symmetric, so the integration over angles is trivial, and that the integrand vanishes outside the region $0<\tau<\delta\tau$. 
Notice as well that we have
changed from the global time coordinate to proper time $\tau$, which accounts for the Jacobian factor, where we have
defined
\begin{eqnarray}
f(\tau)\equiv 1-{2G_{N}M\over R(\tau)}.
\end{eqnarray}

Substituting the expression of the perturbation given in Eq. \eqref{eq:form_of_perturbation}, we can carry out the integration over
the radial coordinate, to find
\begin{eqnarray}
\phi(x)&=&{\delta\tau\sigma_{0}R_{0}^{2}\over M_{\rm Pl}}
\int_{-\infty}^{\infty}{\omega d\omega\over 2\pi}e^{-i\omega t}\int_{0}^{\delta \tau} d\tau\,
e^{i\omega \tau}{\dot{R}(\tau)\over R(\tau)^{3}}{\sqrt{\dot{R}(\tau)^{2}+f(\tau)}\over f(\tau)}
\label{phi(x)_int1} \\[0.2cm]
&\times& \Bigg[
2R(\tau)^{2}\widetilde{g}_{0}\Big(\omega r,\omega R(\tau)\Big)
-2R(\tau)^{2}\widetilde{g}_{0}\Big(\omega r,\omega R(\tau)\Big)-\omega 
R(\tau)^{3}\partial_{2}\widetilde{g}_{0}\Big(\omega r,\omega R(\tau)\Big)
\Bigg] \nonumber \\[0.2cm]
&=&-{\delta\tau\sigma_{0}R_{0}^{2}\over M_{\rm Pl}}
\int_{-\infty}^{\infty}{\omega^{2} d\omega\over 2\pi}e^{-i\omega t}\int_{0}^{\delta \tau} d\tau\,
e^{i\omega \tau}\dot{R}(\tau){\sqrt{\dot{R}(\tau)^{2}+f(\tau)}\over f(\tau)}
\partial_{2}\widetilde{g}_{0}\Big(\omega r,\omega R(\tau)\Big),
\nonumber
\end{eqnarray}
where $\partial_{2}$ indicates the derivative with respect to the second argument of the function. 

As explained above, we have to assume that the radius of the shell is located well below the Vainshtein radius 
$(R_{0}\ll r_{\star})$, whereas we are interested
in the radiation reaching an observer located far away from the source $(r\gg r_{\star})$. 
We are therefore in the so-called radiation limit, 
$\xi'\ll \omega r_{\star} \ll \xi$,
where the
function $\widetilde{g}_{0}(\xi,\xi')$ takes the form \cite{chu_trodden} 
\begin{eqnarray}
\widetilde{g}_{0}(\xi,\xi')=h^{(1)}_{0}(\xi)C_{0}^{\rm (rad)}(\omega r_{\star})\xi'{}^{1\over 4}J_{-{1\over 4}}\left({\sqrt{3}\over 2}\xi'
\right),
\label{eq:g_0_xi_xi'}
\end{eqnarray}
with $h^{(1)}_{0}(\xi)=(i\xi)^{-1}e^{i\xi}$ the zeroth-order spherical Hankel function of the first kind. The explicit form of 
$C_{0}^{(\rm rad)}(\omega r_{\star})$ depends on the frequency regime
\begin{eqnarray}
C_{0}^{(\rm rad)}(\omega r_{\star})=
\left\{
\begin{array}{ll}
{1\over (\omega r_{\star})^{3/4}}\sqrt{\pi\over 2}e^{{7\pi i\over 8}-i\omega r_{\star}I_{\infty}} & |\omega r_{\star}|\gg 1 \\[0.2cm]
i{3^{1\over 8}\pi\over \Gamma\left({1\over 4}\right)} & |\omega r_{\star}|\ll 1
\end{array}
\right.,
\label{eq:asymp_C0_rad}
\end{eqnarray}
where $I_{\infty}\approx 0.253$. 

In order to compute the derivative in the integrand of Eq. \eqref{phi(x)_int1}, we use the Bessel function recursion relation
$[z^{-\nu}J_{\nu}(z)]'=-z^{-\nu}J_{\nu+1}(z)$. In addition, the integral admits a further simplifications in the case of a
nonrelativistic collapse: 
taking the radius of the shell much larger than its Schwarzschild radius we can set $f(\tau)\approx 1$, whereas asuming
its velocity during the collapse process to be much smaller than the speed of light we have $\dot{R}(\tau)\ll 1$. With this, we arrive at
\begin{eqnarray}
\phi(x)&=&-{i\sqrt{3}\sigma_{0}R_{0}^{2}\delta\tau\over 4\pi M_{\rm Pl}}{1\over r}\int_{0}^{\delta \tau}d\tau\,\dot{R}(\tau)R(\tau)^{1\over 4}
\nonumber \\[0.2cm]
& & \times\,\,
\int_{-\infty}^{\infty}d\omega\,\omega^{5\over 4}e^{-i\omega(t-r-\tau)}C_{0}^{(\rm rad)}(\omega r_{\star})J_{3\over 4}\left({\sqrt{3}\over 2}
\omega R(\tau)\right).
\end{eqnarray} 
For a very short implosion, the integral over $\tau$ can be linearized to find
\begin{eqnarray}
\phi(x)=-{i\sqrt{3}\sigma_{0}R_{0}^{9\over 4}\dot{R}_{0}\delta\tau^{2}\over 4\pi M_{\rm Pl}}{1\over r}
\int_{-\infty}^{\infty}d\omega\,\omega^{5\over 4}e^{-i\omega(t-r)}C_{0}^{(\rm rad)}(\omega r_{\star})J_{3\over 4}\left({\sqrt{3}\over 2}
\omega R_{0}\right). 
\label{eq:phi_simplified1}
\end{eqnarray}
We split now the integral into two pieces, and use $J_{\nu}(-x)=(-1)^{\nu}J_{\nu}(x)$ together
with the identity 
\begin{eqnarray}
C_{0}^{(\rm rad)}(-\omega r_{\star})=-C_{0}^{(\rm rad)}(\omega r_{\star})^{*},
\end{eqnarray} 
which follows 
from Eqs. \eqref{eq:Green_func_full} and \eqref{eq:g_0_xi_xi'}. After a few manipulations, 
we can write the integral 
as
\begin{eqnarray}
\phi(x)={\sqrt{3}\sigma_{0}R_{0}^{9\over 4}\dot{R}_{0}\delta\tau^{2}\over 2\pi M_{\rm Pl}}{1\over r}
{\rm Im\,}\int_{0}^{\infty}d\omega\,\omega^{5\over 4}e^{-i\omega(t-r)}C_{0}^{(\rm rad)}(\omega r_{\star})J_{3\over 4}\left({\sqrt{3}\over 2}
\omega R_{0}\right). 
\label{eq:phi_simplified2}
\end{eqnarray}

Since we do not have an expression for $C_{0}^{(\rm rad)}(\omega r_{\star})$ 
valid in the whole range of frequencies we estimate the integral by splitting the integration range into a 
high frequency ($\omega \gg \omega_{0}$) and a low frequency ($\omega \ll \omega_{0}$) piece where we substitute the two asymptotic expressions in \eqref{eq:asymp_C0_rad}. On general grounds we can assume that the matching
takes place at a frequency $\omega_{0} r_{\star}\sim 1$
\begin{eqnarray}
\int_{0}^{\infty}d\omega\,\omega^{5\over 4}e^{i\omega(r-t)}C_{0}^{(\rm rad)}(\omega r_{\star})J_{3\over 4}\left({\sqrt{3}\over 2}
\omega R_{0}\right) \approx
{3^{1\over 8}\pi\over\Gamma\left({1\over 4}\right)}
\int_{0}^{1\over r_{\star}}d\omega\,\omega^{5\over 4}e^{-i\omega(t-r)}J_{3\over 4}\left({\sqrt{3}\over 2}
\omega R_{0}\right)
\nonumber \\[0.2cm]
+{e^{7\pi i\over 8}\over r_{\star}^{3/4}}{\sqrt{\pi\over 2}}
\int_{1\over r_{\star}}^{\infty}d\omega\,\omega^{1\over 2}e^{-i\omega(t-r+r_{\star}I_{\infty})}J_{3\over 4}\left({\sqrt{3}\over 2}
\omega R_{0}\right).
\hspace*{2cm}
\end{eqnarray}
For the first integral, we find in the limit $R_{0}\ll r_{\star}$
\begin{eqnarray}
{3^{1\over 8}\pi\over\Gamma\left({1\over 4}\right)}
\int_{0}^{1\over r_{\star}}d\omega\,\omega^{5\over 4}e^{-i\omega(t-r)}J_{3\over 4}\left({\sqrt{3}\over 2}
\omega R_{0}\right)&=&{3^{1\over 8}\pi\over\Gamma\left({1\over 4}\right)}{1\over R_{0}^{9/4}}
\int_{0}^{R_{0}/r_{\star}}dx\,x^{5\over 4}e^{-ix{t-r\over R_{0}}}J_{3\over 4}\left({\sqrt{3}\over 2}x\right)
\nonumber \\[0.2cm]
&\approx&{1\over 3\sqrt{3}R_{0}^{9/4}}\left({R_{0}\over r_{\star}}\right)^{3}.
\label{eq:int_field_lowf}
\end{eqnarray}
A similar change of variables in the second integral gives, in the same regime, the result
\begin{eqnarray}
{e^{7\pi i\over 8}\over r_{\star}^{3/4}}{\sqrt{\pi\over 2}}
\int_{1\over r_{\star}}^{\infty}d\omega\,\omega^{1\over 2}e^{-i\omega(t-r+r_{\star}I_{\infty})}J_{3\over 4}\left({\sqrt{3}\over 2}
\omega R_{0}\right)\hspace*{3cm} \nonumber \\[0.2cm]
\approx\sqrt{\pi\over 2}{e^{7\pi i\over 8}\over r_{\star}^{3/4}R_{0}^{3/2}}\int_{0}^{\infty}dx\, 
x^{1\over 2}e^{-ix{t-r+r_{\star}I_{\infty}\over 
R_{0}}}J_{3\over 4}\left({\sqrt{3}\over 2}x\right).
\label{eq:int_field_highf}
\end{eqnarray}

The expression of the field given in \eqref{eq:phi_simplified2} shows that both the low and high frequency contributions
to the integral come multiplied by an overall factor $R_{0}^{9/4}$. This means that the low frequency contribution shown in Eq. 
\eqref{eq:int_field_lowf} is suppressed by a factor $(R_{0}/r_{\star})^{3}$, whereas the prefactor of the high frequency modes
is just $(R_{0}/r_{\star})^{3/4}$. As a consequence, due to the relative suppression of the low with respect to the
high frequency modes we can neglect the former and write
\begin{eqnarray}
\phi(t,r)\approx \sqrt{3\over 8\pi}{\sigma_{0}\dot{R}_{0}\delta\tau^{2}\over M_{\rm Pl}}\left({R_{0}\over r_{\star}}\right)^{3\over 4}
{1\over r}{\rm Im}\int_{0}^{\infty}dx\, 
x^{1\over 2}e^{-ix{t-r+r_{\star}I_{\infty}\over 
R_{0}}+{7\pi i\over 8}}J_{3\over 4}\left({\sqrt{3}\over 2}x\right).
\label{eq:solution_phi_galileon}
\end{eqnarray}

From Eq. \eqref{eq:asymp_C0_rad}, and given the asymptotic behavior of the Bessel function for large values of the argument, we find 
that the integrand in this expression is not damped at high frequencies 
$|\omega|\rightarrow\infty$, but oscillating. 
This feature of the solution is an artefact of our choice of zero width distribution \eqref{eq:form_of_perturbation} for the source in 
Eq. \eqref{eq:laplacian_eff_metric}. Not only the infinitely thin source pumps in energy 
at all scales, including the transplanckian region, but due to the presence of a $\delta'(x)$ function in the source, there
is an enhancement of the contribution of higher frequencies. 
This is the origin of the absence of a characteristic damping scale for the frequencies in Eq. \eqref{phi(x)_int1}.

This is indeed a problem from the point of view that we are dealing with an effective field theory valid below some energy
scale $\Lambda$. A physical way to avoid this is to consider a finite size source, in such a way that gradients
in the Galileon field are kept below $\Lambda^{2}$. This indeed makes the analysis much more involved. 
Here we use an alternative procedure consisting in introducing a physical cutoff function in the
integral suppressing high frequencies. The scale of the cutoff is determined by the characteristic width of the collapsing shell
$\Delta$ which is also bound by the cutoff scale, $\Delta\gg \Lambda^{-1}$.
In the following, we use an exponential damping factor $e^{-\epsilon x}$, where $\epsilon$ will be taken to be of
the order
\begin{eqnarray}
\epsilon\sim {\Delta\over R_{0}}\ll 1,
\label{eq:physical_cutoff}
\end{eqnarray}
As it will be seen later, 
other choices of the cutoff function lead to modifications of our result by factors of
order one. 
Thus, our analysis is valid in the regime
\begin{eqnarray}
\Lambda^{-1}\ll \Delta \ll R_{0} \ll r_{\star}\ll r.
\end{eqnarray} 
This in particular means that the radius of the collapsing shell should satisfy $R_{0}\gg (\epsilon\Lambda)^{-1}$.
 
\begin{figure}[t]
\centerline{\includegraphics[scale=0.67]{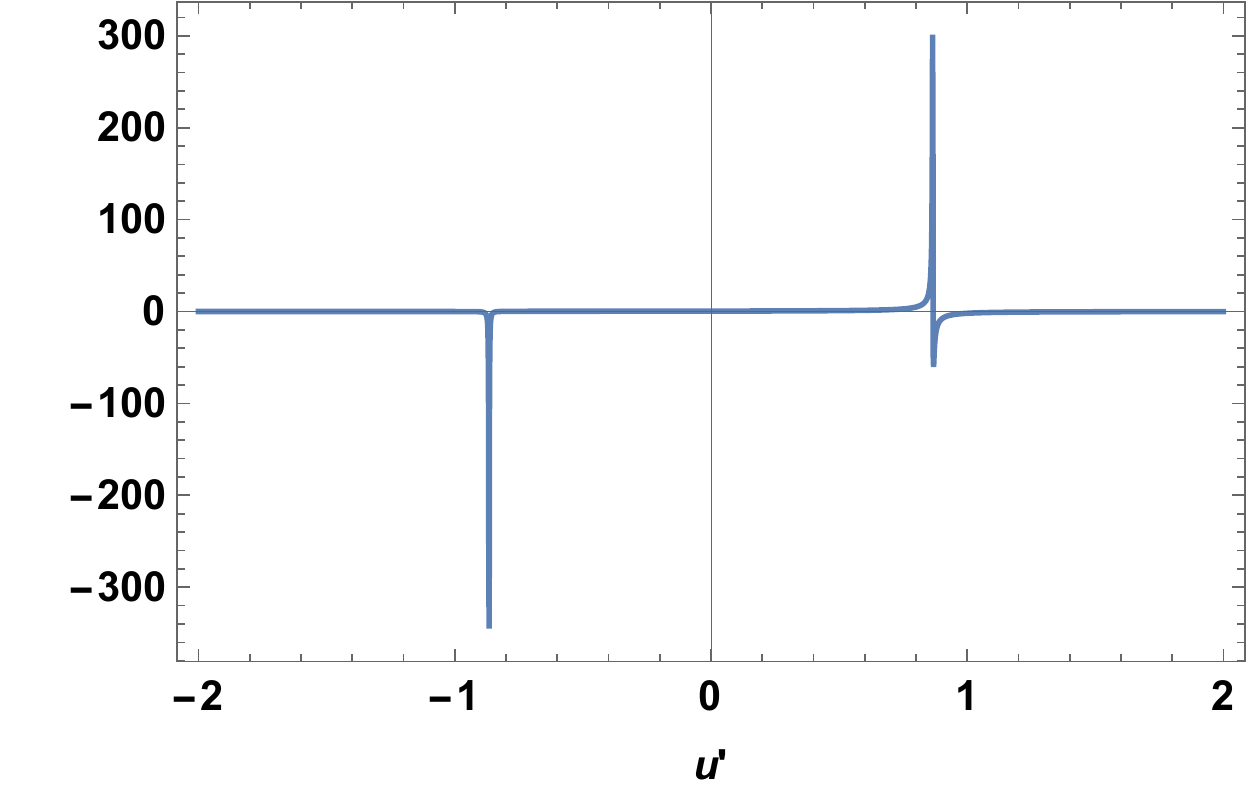}}
\caption[]{Profile of the imaginary part of the integral in Eq. \eqref{eq:solution_phi_galileon} as a function of 
$u'=(t-r+r_{\star}I_{\infty})/R_{0}$ with a cutoff $\epsilon=0.001$.}
\label{fig:field_profile}
\end{figure}

The imaginary part of the integral in \eqref{eq:solution_phi_galileon} can be computed numerically. To optimize the calculation 
we use an adaptive mesh in which, after a first sample, the density of points increases in those regions with a finer structure.
The results are shown in Fig. \ref{fig:field_profile}. We see that a distant observer located at $r>r_{\star}$ (the regime of validity of our analysis) observes
two consecutive pulses in the profile of the Galileon field centered at
\begin{eqnarray}
t-r=-I_{\infty}r_{\star}\pm {\sqrt{3}\over 2}R_{0},
\label{eq:pulse_location}
\end{eqnarray}
where the time difference between the two flashes only depends on the radius of the shell. 
For $R_{0}\gg 2G_{N}M$, a light ray emitted from the surface of the collapsing shell at $t=0$ propagates along $t-r\approx 0$.
Since we are assuming that $R_{0}\ll r_{\star}$, we find that both Galileon pulses will arrive before the light ray by a 
time interval
\begin{eqnarray}
\Delta t=-I_{\infty}r_{\star}\pm {\sqrt{3}\over 2}R_{0}\approx 
-0.253\left({4\sigma_{0} R_{0}^{2}\over M_{\rm Pl}\Lambda^{3}}\right)^{1\over 3},
\end{eqnarray}
where the time difference between the pulses is very small compared to the time of arrival. 
As a consequence, we find that the Galileon field pulses travel ahead of the light fronts.

\section{Energy radiation}
\label{sec:energy}

Next we evaluate the energy radiated during the implosion. 
Computing the energy-momentum tensor for the Galileon 
perturbation, 
the energy radiated by solid angle is given by \cite{chu_trodden}
\begin{eqnarray}
{dE\over d\Omega}=-\lim_{r\rightarrow\infty}\int_{-\infty}^{\infty}dt\,r^{2}\partial_{t}\phi(x)\partial_{r}\phi(x).
\end{eqnarray} 
To evaluate the integrand in this expression, 
we notice that in the solution given in Eq. \eqref{eq:solution_phi_galileon}
all dependence on $t$ and $r$ comes through the combination $t-r$, apart from the overall $1/r$ factor. This leads to the following
relation between the time and radial derivatives 
\begin{eqnarray}
\partial_{r}\phi(x)=-\partial_{t}\phi(x)+\mathcal{O}\left({1\over r^{2}}\right).
\end{eqnarray}
At large distances we can neglect the $r^{-2}$ corrections and write
\begin{eqnarray}
{dE\over d\Omega}\approx {1\over R_{0}}\int_{-\infty}^{\infty}du'\,\{\partial_{u'}[r\phi(u',r)]\}^{2},
\end{eqnarray}
where $u'=(t-r+r_{\star}I_{\infty})/R_{0}$ and the right-hand side is independent of $r$. As expected from the symmetry of the problem, 
energy emission is isotropic.

\begin{figure}
\centerline{\includegraphics[scale=0.67]{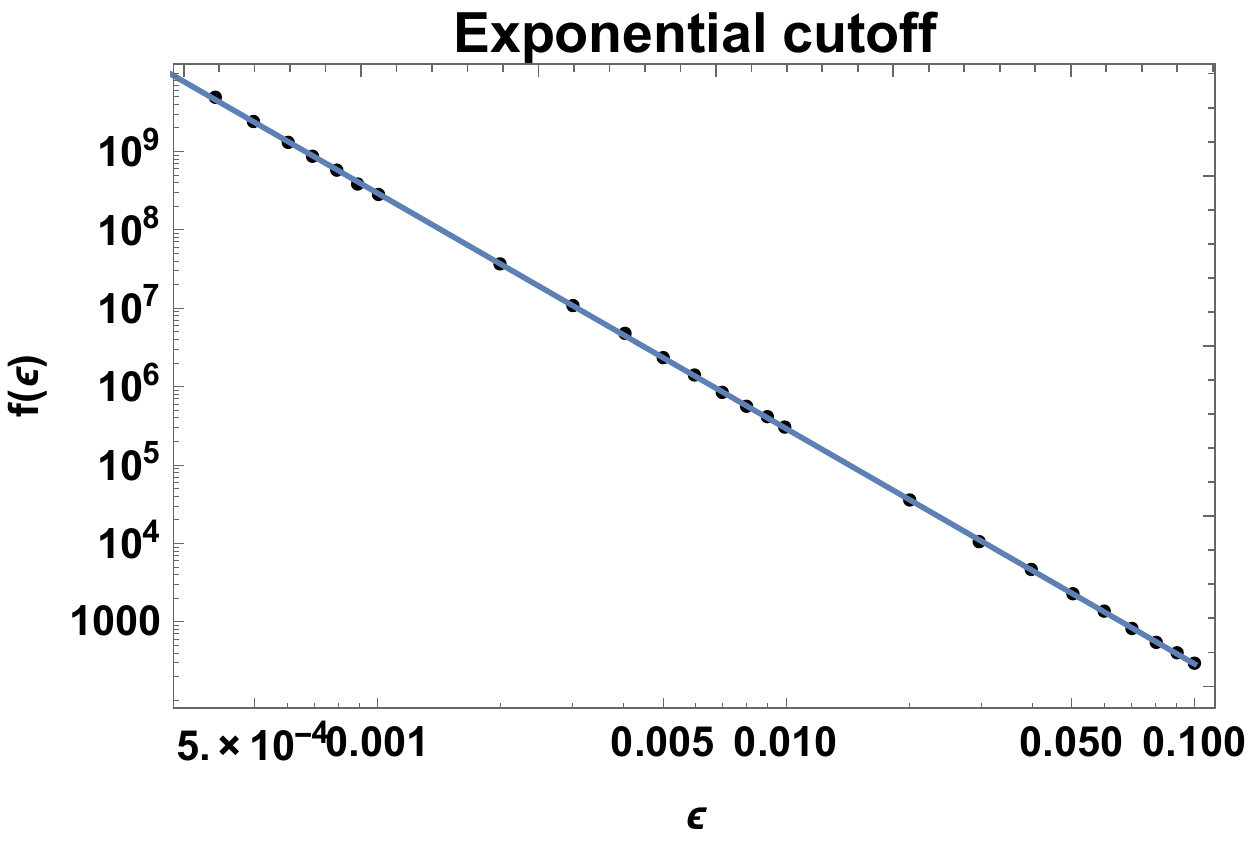}\hspace*{1cm}\includegraphics[scale=0.67]{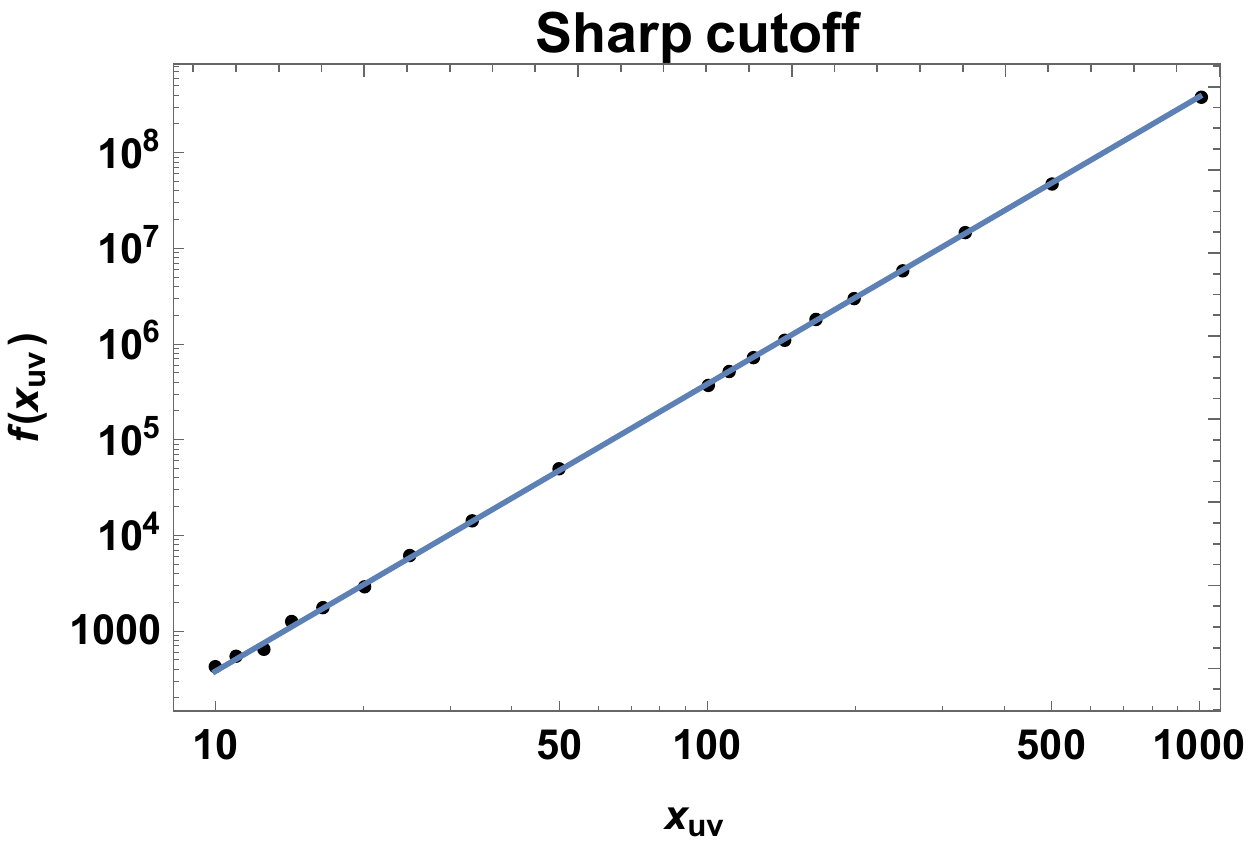}}
\caption[]{In the left panel we see the 
logarithmic plot of the numerical results for the function $f(\epsilon)$ for $\epsilon$ in the range $[0.0005,0.1]$.
The solid line represents the fit in Eq. \eqref{eq:f(epsilon)fit}. The right panel shows the corresponding plot for a sharp 
cutoff $10\leq x_{\rm uv}\leq 1000$ with the fit \eqref{eq:f(xuv)fit}.}
\label{fig:loglogplot}
\end{figure}

The numerical solution for the Galileon field shown in Fig. \ref{fig:field_profile} indicates that the integrand
is strongly damped for large values or $|u'|$, which guarantees the convergence of the total energy radiated during the
collapse, the contribution to the integral being peaked around the two pulses at $u'=\pm \sqrt{3}/2$. The result
can be written as
\begin{eqnarray}
{dE\over d\Omega}\approx {3\sigma_{0}^{2}\dot{R}_{0}^{2}\delta\tau^{4}\over 8\pi M_{\rm Pl}^{2}R_{0}}\left({R_{0}\over r_{\star}}\right)^{3\over 2}
f(\epsilon),
\label{eq:dE/dOmega}
\end{eqnarray}
where $f(\epsilon)$ is a function of the physical cutoff scale \eqref{eq:physical_cutoff} that can be computed 
numerically from the solution $\phi(u',r)$, using again an adaptive mesh to evaluate the integral. 

Proceeding in this way, the results show that $f(\epsilon)$ has a simple scaling with $\epsilon$, as shown by the
logarithmic plot in the left panel of Fig. \ref{fig:loglogplot}. A fit of the numerical solution leads to the solution
\begin{eqnarray}
f(\epsilon)=0.286\, \epsilon^{-3.003}.
\label{eq:f(epsilon)fit}
\end{eqnarray}
Increasing the precision in the evaluation of the integral, i.e. the density of points
around the peaks at $u=\pm\sqrt{3}/2$, shows that the exponent consistently approaches 3. 
Moreover, it can be checked that other choices for the cutoff do not seem to modify this exponent. 
For example, solving the integral in Eq. \eqref{eq:solution_phi_galileon} with a sharp frequency cutoff 
$x_{\rm uv}\sim R_{0}/\Delta$ leads to the following scaling for the function in Eq. \eqref{eq:dE/dOmega}
\begin{eqnarray}
f\left(x_{\rm uv}\right)=0.371 x_{\rm uv}^{3.005}.
\label{eq:f(xuv)fit}
\end{eqnarray}  
The right panel of Fig. \ref{fig:loglogplot} shows a logarithmic plot of the numerical results in this case, 
together with the previous fit function.

From this
we infer the following expression for the total energy emitted by the collapsing shell to be
\begin{eqnarray}
E=\mathcal{C}{\sigma_{0}^{2}\dot{R}_{0}^{2}\delta\tau^{4}\over M_{\rm Pl}^{2}R_{0}}\left({R_{0}^{3}\over r_{\star}\Delta^{2}}
\right)^{3\over 2},
\label{eq:total_energy_final}
\end{eqnarray}
where $\mathcal{C}$ is a numerical constant of order 1 depending upon the details of the collapsing object.
We see how the overall size of the total energy radiated results from the competition of two effects: 
a Vainshtein suppression by a factor $(R_{0}/r_{\star})^{3/2}$ and the enhancement due to the finite width effects scaling
as $(R_{0}/\Delta)^{3}$. This contrasts with what is found for the Galileon radiation from a binary system, where the 
suppression factor is not determined by the characteristic size of the system but by its frequency \cite{dRMT1}.

It is important to stress that the dependence on $\Delta$ in Eq. \eqref{eq:total_energy_final} cannot be considered
a spurious effect. On physical grounds, it is expected that quantities such as the radiated energy depend on the details of
the shell, in particular its effective width. In our approach this width is introduced as a physical scale cutting the contributions of
high frequencies off. Being a physical cutoff, there is every reason to expect that the final result keeps a memory of it. Moreover, 
we have seen that the scaling of $\Delta$ is robust with respect to different mathematical  implemententations of the cutoff scale. 

\section{The next-to-leading order}
\label{sec:nlo}

Our previous analysis was made under the assumption that the implosion of the shell occurs with nonzero initial velocity, $\dot{R}_{0}\neq 0$.
In order to consider the collapse from rest rather than an implosion, 
we need to compute the perturbation of the trace of the energy-momentum tensor at second
order in $\delta \tau$. 
\begin{eqnarray}
\delta T&=&{\delta\tau \dot{R}(\tau)\over R(\tau)}\sigma(\tau)\left[R(\tau)\delta'\Big(r-R(\tau)\Big)+2
\delta\Big(r-R(\tau)\Big)\right]
\nonumber \\[0.2cm]
&-&{\delta\tau^{2}\dot{R}(\tau)^{2}\over R(\tau)^{2}}\sigma(\tau)\left[3\delta\Big(r-R(\tau)\Big)
+2R(\tau)\delta'\Big(r-R(\tau)\Big)+{1\over 2}\delta''\Big(r-R(\tau)\Big)\right]  \\[0.2cm]
&+&{\delta\tau^{2}\ddot{R}(\tau)\over R(\tau)}\sigma(\tau)
\left[\delta\Big(r-R(\tau)\Big)+{1\over 2}R(\tau)\delta'\Big(r-R(\tau)\Big)
\right].
\nonumber
\end{eqnarray}
In order to preserve the pertubative expansion, we impose the ``slow roll'' conditions
\begin{eqnarray}
{\delta\tau\dot{R}(\tau)\over R(\tau)}\ll 1, \hspace*{1cm}
{\delta\tau^{2}\dot{R}(\tau)^{2}\over R(\tau)^{2}}\sim
{\delta\tau^{2}\ddot{R}(\tau)\over R(\tau)}\ll 1.
\end{eqnarray}
Adding the next-to-leading order correction to the source in Eq. \eqref{eq:laplacian_eff_metric} allows for a resolution of 
the Galileon perturbation in the form
\begin{eqnarray}
\phi(x)=[\phi(x)]_{1}+[\phi(x)]_{2},
\end{eqnarray}
where $[\phi(x)]_{1}$ is the solution found in the Eq. \eqref{eq:solution_phi_galileon}.

With these expressions we can calculate the next-to-leading order corrections to the Galileon radiation process studied in 
previous sections. It can be seen that this correction has a structure similar to the leading term. Again we find two pulses 
located at the positions given in Eq. \eqref{eq:pulse_location}. 
Here, however, we will be interested instead in the case of a matter shell at the onset of gravitational collapse
from rest, for which the leading order contribution vanishes, $[\phi(x)]_{1}=0$. 
Setting $\dot{R}_{0}=0$ and following the same steps and approximations as in Sec. \ref{sec:pert}, 
we arrive at the following expression for
the perturbation of the Galileon field
\begin{eqnarray}
\phi(x)&=&-{ R_{0}^{2}\ddot{R}_{0}\sigma_{0}\delta\tau^{3}\over M_{\rm Pl}}\int_{-\infty}^{\infty}{\omega^{2}d\omega\over 2\pi}
e^{-i\omega t}\partial_{2}\widetilde{g}_{0}(\omega r,\omega R_{0})
\nonumber \\[0.2cm]
&=&-{i\sqrt{3}R_{0}^{2}\ddot{R}_{0}\sigma_{0}\delta\tau^{3}\over 4\pi M_{\rm Pl}}{1\over r}
\int_{-\infty}^{\infty} d\omega \omega^{5\over 4}e^{-i\omega(t-r)}C_{0}^{(\rm rad)}(\omega r_{\star})J_{3\over 4}\left(
{\sqrt{3}\over 2}\omega R_{0}\right).
\end{eqnarray}
Comparing this result with the one found in Eq. \eqref{eq:phi_simplified1}, we see that the only difference with respect to that
calculation presented in the previous section is that now we have a different prefactor depending on $\ddot{R}_{0}$ rather than $\dot{R}_{0}$. 
Physically, we find the same profile for the Galileon field depicted in Fig. \ref{fig:field_profile}, two successive pulses
travelling ahead of the light front. As for the total energy radiated, we find 
\begin{eqnarray}
E=\mathcal{C}'{\sigma_{0}^{2}\ddot{R}_{0}^{2}\delta\tau^{6}\over M_{\rm Pl}^{2}R_{0}}\left({R_{0}^{3}\over r_{\star}\Delta^{2}}
\right)^{3\over 2},
\end{eqnarray}
where again $\mathcal{C}'$ is a numerical constant of order one.

\section{Closing remarks}
\label{eq:conc}

Apart from their intrinsic interest in classical and quantum field theory, 
Galileons emerge in theories of massive gravity and therefore provide
a window to test alternative theories of gravity based on deformations of the Einstein-Hilbert action by relevant operators. 
In particular, astrophysics may provide a number of physical scenarios where Galileon theories could be put to the test. 
Here we have presented a tentative study of the problem of Galileon radiation in spherical gravitational collapse. Choosing 
spherical symmetry has two consequences: it simplifies the problem from a technical point of view and also eliminates the GR background radiation leaving a distinct Galileon signal. 
Although quite 
simplified, our model could be seen as a first approximation to the problem, 
displaying a number of features expected to be present in more realistic descriptions of astrophysical gravitational collapse. 

Our results indicate the emission of two pulses in the Galileon field traveling at superluminal speed. This is not an unusual 
feature in modified theories of gravity in general \cite{superluminal} and
massive gravity and Galileons in particular \cite{galileon,superluminal_mg}, where nonlinearities may lead to superluminal propagation. 
In the cubic Galileon theory this can be seen from the effective metric \eqref{eq:effective_metric_pert},
whose structure of light cones shows that the phase and group velocity of radial perturbations exceeds the speed of light. 

Galileon theories are known to modify observable effects such as weak lensing \cite{weak_lensing}.
An interesting issue worth considering is the feasibility of direct Galileon field detection. In the theory studied here, the 
Galileon field perturbation couples to the trace of the energy-momentum tensor, unlike ordinary gravitational waves that couple to the transverse-traceless part of the energy-momentum tensor. In both cases, however, their coupling to matter have the same suppression by the Planck scale. Given its superluminal propagation, the Galileon signal should predate the electromagnetic observation of the astrophysical phenomenon sourcing it. Despite the additional Vainshtein suppression, the recent success in the direct detection of gravitational waves 
\cite{LIGO} opens up the possibility of designing experiments sensitive to these extra modes in a 
maybe not-too-distant 
future.

There are various other directions for future work, considering more realistic models of gravitational collapse 
and leaving behind
the approximations used in this paper. One would be using a top-hat window function for the density of the 
collapsing object, i.e. studying the collapse of a homogeneous dust ball instead of the shell considered here. 
At early stages, the Galileon radiation coming from the surface of the object is expected to behave similarly to the 
one produced by the collapsing shell, including the superluminal behavior found above. 
The radiation coming from inner layers, however, would presumably smooth the pulses out into a band profile. 
A full analysis valid for late times would require relaxing some of the approximations used in our analysis. 

Within the context of the cubic Galileon theory, 
it would be interesting to explore the possibility of going beyond the perturbative approach used here. 
This requires solving the full Galileon field equation in the curved background produced by a spherical source. 
Due to the nature of the field equations, this would require the application of more powerful numerical techniques.
Finding such solutions would allow to study the issue of superluminal propagation in a more general fashion.
These and other problems will be addressed elsewhere.

\setcounter{footnote}{0}

\section*{Acknowledgments}

We would like to thank Jos\'e Barb\'on and Kerstin Kunze  
for discussions and comments. J.M.-G. acknowledges partial support from Universidad Aut\'onoma de Madrid. 
The work of M.A.V.-M. has been partially supported by Spanish Government grant FPA2015-64041-C2-2-P. He also thanks  
the Yukawa Institute for Theoretical Physics and Instituto de F\'{\i}sica Te\'orica UAM/CSIC 
for hospitality during the completion of this work.


\begin{thebibliography}{99}

\bibitem{supernovae}
A.~G.~Riess {\it et al.} [Supernova Search Team Collaboration],
  {\it Observational evidence from supernovae for an accelerating universe and a cosmological constant,}
  Astron.\ J.\  {\bf 116} (1998) 1009
{\tt [astro-ph/9805201]}.
\\
 S.~Perlmutter {\it et al.} [Supernova Cosmology Project Collaboration],
  {\it Measurements of $\Omega$ and $\Lambda$ from 42 high redshift supernovae,}
  Astrophys.\ J.\  {\bf 517} (1999) 565
 {\tt [astro-ph/9812133].}

\bibitem{CC}
S.~Weinberg,
  {\it The cosmological constant problems,} in: ``Sources and detection of dark matter and dark energy in the 
  universe", ed. D.~B.~Cline, Springer 2001 
  [{\tt astro-ph/0005265}].
\\
J.~Martin,
  {\it Everything you always wanted to know about the cosmological constant problem (but were afraid to ask),}
  Comptes Rendus Physique {\bf 13} (2012) 566
{\tt  [arXiv:1205.3365 [astro-ph.CO]]}.
\\
A.~Padilla,
  {\it Lectures on the Cosmological Constant Problem,}
  {\tt arXiv:1502.05296 [hep-th]}.


\bibitem{DGP}
G.~R.~Dvali, G.~Gabadadze and M.~Porrati,
  {\it Metastable gravitons and infinite volume extra dimensions,}
  Phys.\ Lett.\ {\bf B484} (2000) 112
  {\tt [hep-th/0002190]}.
\\
G.~R.~Dvali, G.~Gabadadze and M.~Porrati,
  {\it A Comment on brane bending and ghosts in theories with infinite extra dimensions,}
  Phys.\ Lett.\ {\bf B484} (2000) 129
  {\tt [hep-th/0003054]}.
\\
G.~R.~Dvali, G.~Gabadadze and M.~Porrati,
  {\it 4-D gravity on a brane in 5-D Minkowski space,}
  Phys.\ Lett.\ {\bf B485} (2000) 208
  {\tt [hep-th/0005016]}.


\bibitem{lue}
A.~Lue,
  {\it The phenomenology of Dvali-Gabadadze-Porrati cosmologies,}
  Phys.\ Rept.\  {\bf 423} (2006) 1
  {\tt [astro-ph/0510068]}.
\\
C.~Deffayet,
  {\it Theory and phenomenology of DGP gravity,}
  Int.\ J.\ Mod.\ Phys.\ {\bf D16} (2008) 2023.

\bibitem{hinterbichlerRMP}
K.~Hinterbichler,
  {\it Theoretical Aspects of Massive Gravity,}
  Rev.\ Mod.\ Phys.\  {\bf 84} (2012) 671
 {\tt [arXiv:1105.3735 [hep-th]]}.

\bibitem{deRhamLRR}
C.~de Rham,
  {\it Massive Gravity,}
  Living Rev.\ Rel.\  {\bf 17} (2014) 7
  {\tt [arXiv:1401.4173 [hep-th]]}.

\bibitem{degravitation}
G.~Dvali, G.~Gabadadze and M.~Shifman,
  {\it Diluting cosmological constant in infinite volume extra dimensions,}
  Phys.\ Rev.\ D {\bf 67} (2003) 044020
  {\tt [hep-th/0202174]}.
\\
G.~Dvali, S.~Hofmann and J.~Khoury,
  {\it Degravitation of the cosmological constant and graviton width,}
  Phys.\ Rev.\ {\bf D76} (2007) 084006
  {\tt [hep-th/0703027 [HEP-TH]]}.
\\
C.~de Rham, S.~Hofmann, J.~Khoury and A.~J.~Tolley,
  {\it Cascading Gravity and Degravitation,}
  JCAP {\bf 0802} (2008) 011
{\tt  [arXiv:0712.2821 [hep-th]]}.
\\
C.~de Rham,
  {\it Cascading Gravity and Degravitation,}
  Can.\ J.\ Phys.\  {\bf 87} (2009) 201
 {\tt [arXiv:0810.0269 [hep-th]]}.

\bibitem{stueckelberg}
N.~Arkani-Hamed, H.~Georgi and M.~D.~Schwartz,
  {\it Effective field theory for massive gravitons and gravity in theory space,}
  Annals Phys.\  {\bf 305} (2003) 96
  {\tt [hep-th/0210184]}.
M.~D.~Schwartz,
  {\it Constructing gravitational dimensions,}
  Phys.\ Rev.\ {\bf D68} (2003) 024029
 {\tt [hep-th/0303114].}

\bibitem{vainshtein}
A.~I.~Vainshtein,
  {\it To the problem of nonvanishing gravitation mass,}
  Phys.\ Lett.\ {\bf B39} (1972) 393.
\\
E.~Babichev and C.~Deffayet,
  {\it An introduction to the Vainshtein mechanism,}
  Class.\ Quant.\ Grav.\  {\bf 30} (2013) 184001
  {\tt [arXiv:1304.7240 [gr-qc]].}

\bibitem{ostrogradsky}
R.~P.~Woodard,
  {\it Ostrogradsky's theorem on Hamiltonian instability,}
  Scholarpedia {\bf 10} (2015) no.8,  32243
{\tt  [arXiv:1506.02210 [hep-th]]}.
\\
C.~de Rham and A.~Matas,
  {\it Ostrogradsky in Theories with Multiple Fields,}
  {\tt arXiv:1604.08638 [hep-th].}

\bibitem{ghost1}
D.~G.~Boulware and S.~Deser,
  {\it Can gravitation have a finite range?,}
  Phys.\ Rev.\ {\bf D6} (1972) 3368.

\bibitem{dRGT_ghostfree}
C.~de Rham, G.~Gabadadze and A.~J.~Tolley,
  {\it Resummation of Massive Gravity,}
  Phys.\ Rev.\ Lett.\  {\bf 106} (2011) 231101
  {\tt [arXiv:1011.1232 [hep-th]]}.

\bibitem{galileon}
A.~Nicolis, R.~Rattazzi and E.~Trincherini,
  {\it The Galileon as a local modification of gravity,}
  Phys.\ Rev.\ {\bf D79} (2009) 064036
  {\tt  [arXiv:0811.2197 [hep-th]].}

\bibitem{nicolis_rattazzi}
A.~Nicolis and R.~Rattazzi,
  {\it Classical and quantum consistency of the DGP model,}
  JHEP {\bf 0406} (2004) 059
  {\tt  [hep-th/0404159]}.

\bibitem{galileon_reviews}
M.~Trodden and K.~Hinterbichler,
  {\it Generalizing Galileons,}
  Class.\ Quant.\ Grav.\  {\bf 28} (2011) 204003
  {\tt  [arXiv:1104.2088 [hep-th]].}
\\
C.~de Rham,
  {\it Galileons in the Sky,}
  Comptes Rendus Physique {\bf 13} (2012) 666
  {\tt  [arXiv:1204.5492 [astro-ph.CO]].}
\\
   T.~L.~Curtright and D.~B.~Fairlie,
  {\it A Galileon Primer,}
 {\tt arXiv:1212.6972 [hep-th]}.
\\
C.~Deffayet and D.~A.~Steer,
  {\it A formal introduction to Horndeski and Galileon theories and their generalizations,}
  Class.\ Quant.\ Grav.\  {\bf 30} (2013) 214006
  {\tt  [arXiv:1307.2450 [hep-th]].}
\\
A.~Joyce, B.~Jain, J.~Khoury and M.~Trodden,
  {\it Beyond the Cosmological Standard Model,}
  Phys.\ Rept.\  {\bf 568} (2015) 1
  {\tt  [arXiv:1407.0059 [astro-ph.CO]].}
  

\bibitem{oppenheimer-snyder}
J.~R.~Oppenheimer and H.~Snyder,
  {\it On Continued Gravitational Contraction,}
  Phys.\ Rev.\  {\bf 56} (1939) 455.

\bibitem{grav_collapse_revs}
C.~L.~Fryer and K.~C.~B.~New,
  {\it Gravitational waves from gravitational collapse,}
  Living Rev.\ Rel.\  {\bf 14} (2011) 1.
\\
W.~D.~Arnett,
  {\it The Physics Of Gravitational Collapse: An Overview,}
  Annals N.\ Y.\ Acad.\ Sci.\  {\bf 336} (1980) 366.

\bibitem{sexl}
R.~U.~Sexl, {\it Monopole Gravitational Radiation}, 
Phys. Lett. {\bf 20} (1966) 376.

\bibitem{structure_form_papers}
E.~Bellini, N.~Bartolo and S.~Matarrese,
  {\it Spherical Collapse in covariant Galileon theory,}
  JCAP {\bf 1206} (2012) 019
{\tt  [arXiv:1202.2712 [astro-ph.CO]].}
\\
A.~Barreira, B.~Li, C.~M.~Baugh and S.~Pascoli,
  {\it Spherical collapse in Galileon gravity: fifth force solutions, halo mass function and halo bias,}
  JCAP {\bf 1311} (2013) 056
{\tt  [arXiv:1308.3699 [astro-ph.CO]]}.

\bibitem{vaidya_solution}
P.~Rudra and U.~Debnath,
  {\it Gravitational collapse in Vaidya space-time for Galileon gravity theory,}
  Can.\ J.\ Phys.\  {\bf 92} (2014) 1474
 {\tt [arXiv:1402.0350 [physics.gen-ph]].}

\bibitem{dRMT1}
C.~de Rham, A.~Matas and A.~J.~Tolley,
  {\it Galileon radiation from binary systems,}
  Phys.\ Rev.\ {\bf D87} (2013) 064024
{\tt  [arXiv:1212.5212 [hep-th]].}

\bibitem{spherical_galileons}
L.~Berezhiani, G.~Chkareuli, C.~de Rham, G.~Gabadadze and A.~J.~Tolley,
  {\it Mixed Galileons and Spherically Symmetric Solutions,}
  Class.\ Quant.\ Grav.\  {\bf 30} (2013) 184003
{\tt  [arXiv:1305.0271 [hep-th]].}

\bibitem{poisson}
E.~Poisson, {\it A relativist's toolkit: the mathematics of black hole mechanics}, Cambridge 2007.

\bibitem{chu_trodden}
Y.~Z.~Chu and M.~Trodden,
  {\it Retarded Green's function of a Vainshtein system and Galileon waves,}
  Phys.\ Rev.\ {\bf D87} (2013)  024011
 {\tt [arXiv:1210.6651 [astro-ph.CO]]}.

\bibitem{graviton_mass_bounds}
C.~Patrignani {\it et al.} [Particle Data Group],
  {\it Review of Particle Physics,}
  Chin.\ Phys.\ {\bf C40} (2016) 100001.


\bibitem{superluminal}
J.~\O.~Lindroos, C.~Llinares and D.~F.~Mota,
  {\it Wave Propagation in Modified Gravity,}
  Phys.\ Rev.\ {\bf D93} (2016) no.4,  044050
{\tt  [arXiv:1512.00615 [gr-qc]].}

\bibitem{superluminal_mg}
A.~Adams, N.~Arkani-Hamed, S.~Dubovsky, A.~Nicolis and R.~Rattazzi,
  {\it Causality, analyticity and an IR obstruction to UV completion,}
  JHEP {\bf 0610} (2006) 
  {\tt  [hep-th/0602178].}
\\
G.~L.~Goon, K.~Hinterbichler and M.~Trodden,
  Phys.\ Rev.\ {\bf D83} (2011) 085015
  doi:10.1103/PhysRevD.83.085015
  [arXiv:1008.4580 [hep-th]].
\\
P.~de Fromont, C.~de Rham, L.~Heisenberg and A.~Matas,
  {\it Superluminality in the Bi- and Multi- Galileon,}
  JHEP {\bf 1307} (2013) 067
 {\tt  [arXiv:1303.0274 [hep-th]]}.
\\
S.~Garcia-Saenz,
  {\it Behavior of perturbations on spherically symmetric backgrounds in multi-Galileon theory,}
  Phys.\ Rev.\ {\bf D87} (2013) no.10,  104012
{\tt  [arXiv:1303.2905 [hep-th]].}
\\
C.~De Rham, L.~Keltner and A.~J.~Tolley,
  {\it Generalized galileon duality,}
  Phys.\ Rev.\ {\bf D90} (2014) 024050
{\tt  [arXiv:1403.3690 [hep-th]].}
\\
 R.~Kolevatov,
  {\it Superluminality in dilatationally invariant generalized Galileon theories,}
  Phys.\ Rev.\ {\bf D92} (2015)  123532
 {\tt  [arXiv:1508.00046 [hep-th]].}

\bibitem{weak_lensing}
Y.~Park and M.~Wyman,
  {\it Detectability of Weak Lensing Modifications under Galileon Theories,}
  Phys.\ Rev.\ {\bf D91} (2015)  064012
 {\tt  [arXiv:1408.4773 [astro-ph.CO]].}
  
\bibitem{LIGO}
B.~P.~Abbott {\it et al.} [LIGO Scientific and Virgo Collaborations],
  {\it Observation of Gravitational Waves from a Binary Black Hole Merger,}
  Phys.\ Rev.\ Lett.\  {\bf 116} (2016)  061102
{\tt  [arXiv:1602.03837 [gr-qc]].}


\end{thebibliography}
\end{document}